\documentclass{pss}
\usepackage{epsf,pstricks}
\usepackage{amsmath}
\usepackage{epsfig,graphicx}
\begin{document}

\title{Spin triplet superconductivity in Sr$_2$RuO$_4$}
\author{Karol I. Wysoki\'nski$^1$,
G. Litak$^2$, J.F. Annett$^3$, B.L. Gy\"{o}rffy$^3$}
\address{$^1$Institute of Physics, M. Curie -- Sk\l{}odowska University,
 ul. Radziszewskiego 10, 20-031 Lublin, Poland
\\ $^2$Department of Mechanics, Technical University of Lublin,
Nadbystrzycka 36,
20-618 Lublin, Poland \\
$^3$H.H. Wills Physics Laboratory, University of Bristol, Tyndall
Ave, Bristol, BS8 1TL, UK}

\submitted{May 06, 2002}
\maketitle

\hspace{9mm}Subject classification:74.70.Pq,74.20.Rp,74.25.Bt

\begin{abstract}
 Sr$_2$RuO$_4$ is at present the best candidate for being a
superconducting analogue of the triplet superfluidity in $^3$He.
This material is a good (albeit correlated) Fermi liquid in the normal
state and an exotic superconductor below $T_c$. The  mechanism of
superconductivity  and symmetry of the order parameter are the
main puzzling issues of  on-going research. Here we
present the results of our search for a viable description of the
superconducting
state realised in this material. Our calculations are based on
a three-dimensional effective
three-band
 model with a realistic band structure.
We have found a state with non-zero order parameter on each of the
three sheets of the Fermi surface. The corresponding gap in the
quasi-particle spectrum has line or point nodes on the $\alpha$
and $\beta$ sheets and is complex
 with no nodes on the $\gamma$ sheet.
This state describes remarkably well a number of existing
experiments including power low temperature dependence of the
specific heat, penetration depth, thermal conductivity etc. The
stability of the state with respect to disorder and different
interaction parameters are also analyzed briefly.
\end{abstract}
\section{Introduction.}
The discovery of superconductivity in Sr$_2$RuO$_4$
\cite{Maeno_a,Maeno} has generated renewed interest in this
material \cite{Cava}. Originally the main motivation in Ref.
\cite{Maeno_a} was the close similarity of its crystal structure
with that of high temperature superconducting oxide
La$_{2-x}$Ca$_x$CuO$_4$. However, unlike cuprates, strontium
ruthenate is metallic without doping and is not readily
superconducting. Indeed, very pure single crystals have
superconducting transition temperature $T_c=1.5$K, rather low on
the scale of the high T$_c$ cuprates.

In the normal state the electrons in strontium ruthenate form a
well behaved, albeit aniso\-tropic and correlated, Fermi  liquid.
In particular the resistivity shows a typical quadratic dependence
on temperature $\rho_i(T)=\rho_i^0 + A_i T^2$ at low temperatures.
Here $\rho_i^0$ is the `residual'  resistivity measured at
temperature just above $T_c$ and $i$ denotes the direction of the
current ($i=ab$ for in plane  and $i=c$ for c-axis resistivity).
Electronic transport is very anisotropic with the ratio of
$\rho_c/\rho_{ab}$ exceeding 500. Normal state electronic specific
heat shows a typical linear dependence on temperature, $C_e=\gamma
T$, with a high value of $\gamma=37.5$ mJ/K$^2$ mol. The
appreciable enhancement (factor 3-5) of $\gamma$ over the band
structure calculations indicates the existence of strong carrier
correlations in the system. The same conclusion can also be drawn
from the enhanced value of the ratio between Pauli
spin-susceptibility and $\gamma$ (Wilson ratio) and the so called
Kadowaki-Woods ratio of the coefficient $A_{ab}$ to $\gamma$.
These values exceeds those for non-correlated systems.

Consistent with the above picture is the fact that
first-principles calculations based on the Local Density
Approximation (LDA) give a good qualitative account of the
electronic structure \cite{maz}. The energy spectrum around the
Fermi level of strontium ruthenate is dominated by electrons
occupying  $t_{2g}$ orbitals ($d_{xz}$,$d_{yz}$,$d_{xy}$) of
$Ru^{4+}$. The Fermi surface consists of three cylindrical sheets
open in the z-direction.
Those called $\alpha$ and $\beta$ stem from hybridised  $d_{xz}$
and $d_{yz}$ orbitals, while the $\gamma$ sheet is mainly of
$d_{xy}$ character. The $\alpha$ sheet is a hole Fermi surface. It
is centered at the point $X$ of the bct Brillouin zone and
possesses two fold symmetry. The calculated shape of the Fermi
surfaces agrees nicely with that measured {\it via} de Haas--van
Alphen effect \cite{mac}.

While the normal state, as shown above, is quite typical of a
good metal, the superconducting state  is very exotic. The
superconducting transition temperature $T_c$ is strongly
suppressed by even small amount of impurities \cite{miya}. The NMR
measured relaxation  rate shows no Hebel-Slichter peak, and the
$\mu$SR experiments indicate the appearance of spontaneous
magnetic fields at $T<T_c$. Together with a temperature
independent Knight shift, all that points towards very
unconventional, presumably
 spin-triplet, and superconductivity with a time reversal symmetry breaking
pairing state.

Because the electrons forming Cooper pairs in superconductors are
fermions, the pair wave function has to be odd with respect to
interchange. It consists of spin and orbital components. If the
total spin of a pair is zero then such a pairing state is called
spin-singlet. The corresponding spin-wave function is odd so the
orbital part has to be even. In Galilean invariant systems they
correspond to even values of the orbital quantum number
 $l=0,2 \ldots$ and the corresponding superconducting states  are referred to as
having $s-$,$d-$, $\ldots$ wave  symmetry.
  The
 $s-$wave symmetry is realised in  conventional superconductors, while
$d-$wave is the case
 in high temperature superconductors. The electrons in
 strontium ruthenate  most probably pair in relative spin triplet state.
Their spin wave function is even and thus the orbital one has to
be odd. Even though a crystal has only discreet rotational
symmetry, the corresponding symmetries of the superconducting
state are still called $p-$, $f-$, $\ldots$  wave.

\begin{figure}[thb]
\vspace*{-0.5cm}
\centerline{\epsfig{file=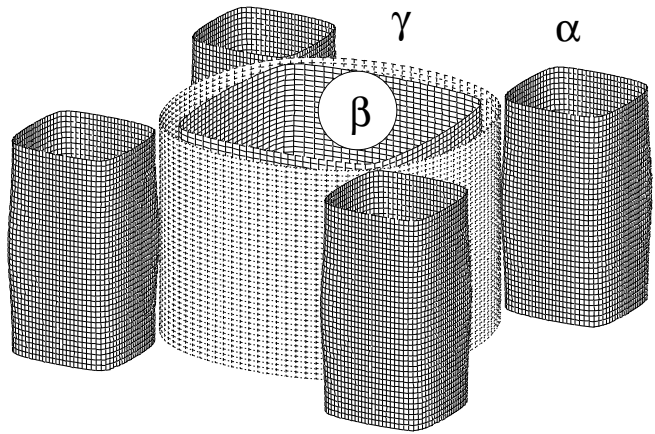,width=10.0cm,angle=-90}}
\vspace{-2.5cm}
\caption{Calculated tight binding Fermi surface of Sr$_2$RuO$_4$.}
\label{fig1}
\end{figure}

As well known, a relatively weak dependence of $T_c$  on the
concentration of impurities and an exponential dependence of the
electron specific heat on temperature point to spin singlet s-wave
superconductivity. On the other hand a power law temperature
dependence of specific heat is characteristic of a superconducting
gap which vanishes
 at points or along lines on the Fermi surface,
and
 a dramatic decrease of $T_c$ with impurity concentration is a signal of
$l > 0$ `wave' paring.

The puzzling behaviour of Sr$_2$RuO$_4$ in the superconducting
state is connected with the facts that NMR, $\mu$SR and related
experiments indicate the realisation of the $\Delta (\vec
k)=e_z(k_x\pm ik_y)$ state which does not vanish at the Fermi
level and that the power law temperature dependence of  the
specific heat \cite{NishiZaki}, penetration depth \cite{DalevH} or
thermal conductivity \cite{thermal-cond} which  require the gap to
vanish along lines of the Fermi surface. The point is that out of
all the symmetry distinct states of a bct crystal \cite{annett}
none of the odd-parity states have to be time reversal symmetry
breaking and also possess gap nodes at the same time ({\it c.f.}
table I). The various states, formally fulfilling both
requirements proposed in literature \cite{graf,Dahm} can be
shown to be the sum of symmetry allowed states with distinct
transition temperatures. The single superconducting transition
observed in all studied samples rules out all these proposals as
the single transition could only be a result of accidental
degeneracy. Instead of relying on a symmetry arguments alone, we
present a methodology which is based on explicit construction of
an effective pairing interaction.

\begin{table}
\caption{\label{tableone} Irreducible representations in a tetragonal
crystal. The symbols $X$, $Y$ $Z$
represent any functions transforming as $x$, $y$ and $z$ under
crystal point group operations, while $I$ represents
any function which is invariant under all point group symmetries.
}
\vspace{1cm}
\hspace{2cm}
\begin{tabular}{|c|c||c|c|}
 \hline
irred. &  basis  & time-reversal   & line nodes \\
repres.      &  functions         & symmetry breaking &
\\
 \hline
 A$_{1g}$ & $I$ & No  & No \\
 A$_{1u}$ & $XYZ(X^2-Y^2)$  & No  & Yes \\
  A$_{2g}$ & $XY(X^2-Y^2)$ & No  & Yes \\
    A$_{2u}$ & $Z$  & No  & Yes
\\
  B$_{1g}$ & $X^2-Y^2$  &   No  & Yes \\
   B$_{1u}$ & $XYZ$   & No  & Yes\\
 B$_{2g}$ & XY & No  & Yes \\
 B$_{2u}$ & $Z(X^2-Y^2)$  & No  & Yes\\
  E$_g$  & $\{XZ,YZ\}$ &  Yes  & Yes \\
  E$_u$  & $\{X,Y\}$  & Yes  & No \\
\hline
\end{tabular}

\end{table}

\begin{figure}[h]
\epsfxsize=7cm
\centerline{\epsfig{file=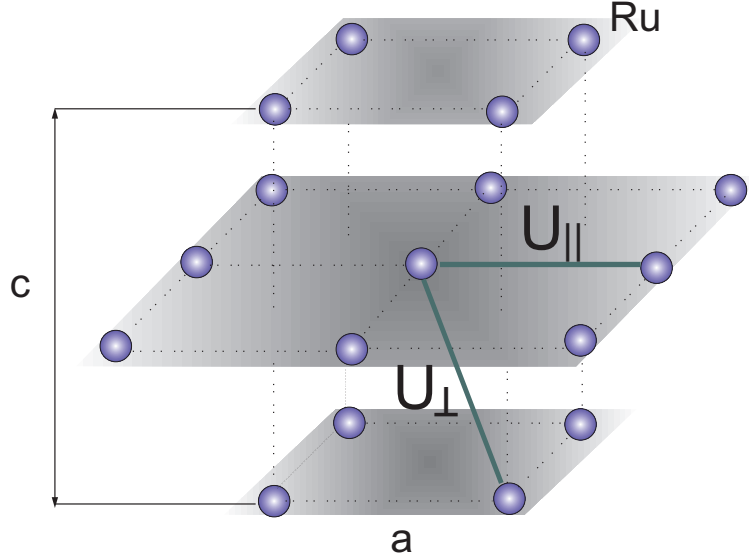,width=10.0cm,angle=0}}
\caption{The relative positions of Ru ions in the tetragonal
$Sr_2RuO_4$ lattice. The  heavy lines symbolise
the main interactions. Interplane $U_{\parallel}$ for the electrons
occupying $d_{xy}$ orbitals at the neighbouring in planes
ions and $U_{\perp}$ for out of plane neighbours and $d_{xz}$ and $d_{yz}$
orbitals.}
\end{figure}

\begin{figure}[h]
\centerline{\epsfig{file=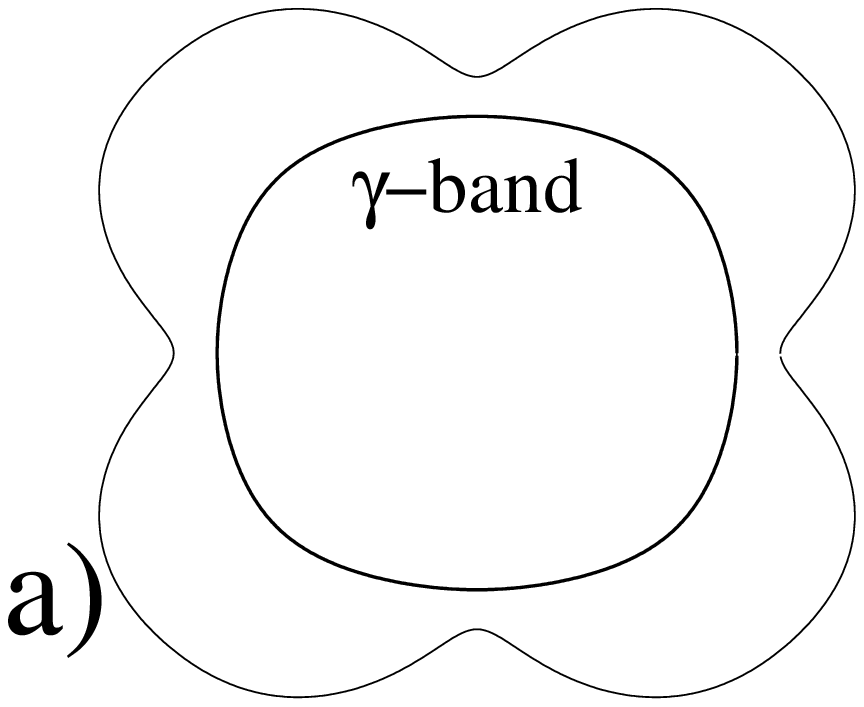,width=5.5cm,angle=0} \hspace{0.5cm}
\epsfig{file=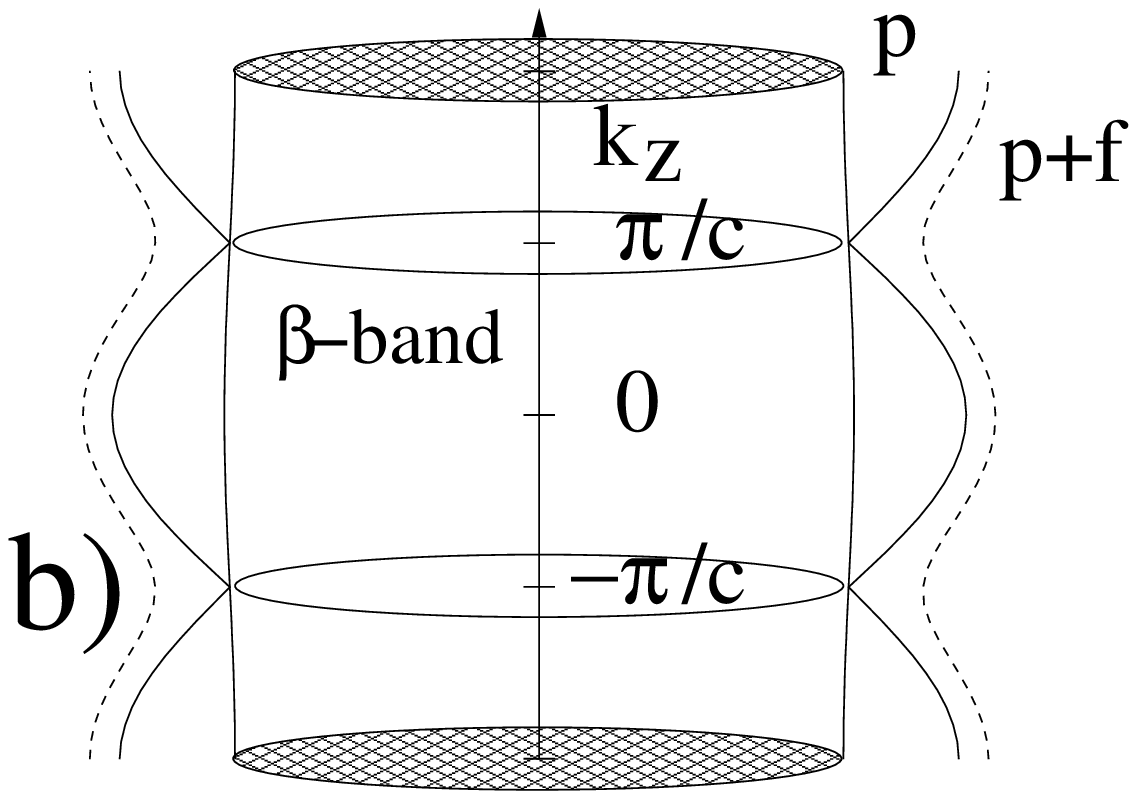,width=6.5cm,angle=0}}
\vspace{0.5cm}
\caption{ Lowest energy
quasiparticle eigenvalues, $E^\nu ({\bf k})$ on the $\gamma$ (a) and 
$\beta$ (b) Fermi
surface sheets; (a) polar plot of
the $\protect\gamma $ sheet in the plane $k_{z}=0$, $E_{\protect%
\gamma ,max}({\bf k}_{F})=0.22{\rm meV}$, $E_{\protect\gamma ,min}({\bf k}
_{F})=0.056{\rm meV}$ (b) vertical cross-section
of the cylindrical $\protect\beta $ sheet in the plane $k_{x}=k_{y}$,
$E_{\protect\beta ,max}({\bf k}_{F})=0.32{\rm meV}$.
A non-zero $f-$wave order
parameter (dotted line) lifts the $p$-wave line nodes (solid lines).}
\label{fig2}
\end{figure}

\begin{figure}[h]
\hspace*{1mm}\centerline{\epsfig{file=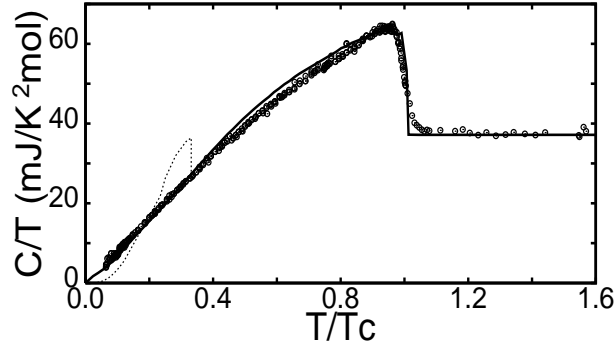,width=4.0cm,angle=-90}}
\vspace{1cm} \caption{Calculated specific heat, $C$, as a function
of temperature, $T$, compared to the experimental data of
NishiZaki {\it et al.}\cite{NishiZaki}. The dotted-line includes
the $f$-wave order parameter, the solid line is without $f$-wave.}
\label{specific}
\end{figure}

 \begin{figure}[h]
\epsfxsize=7cm
\centerline{\epsfig{file=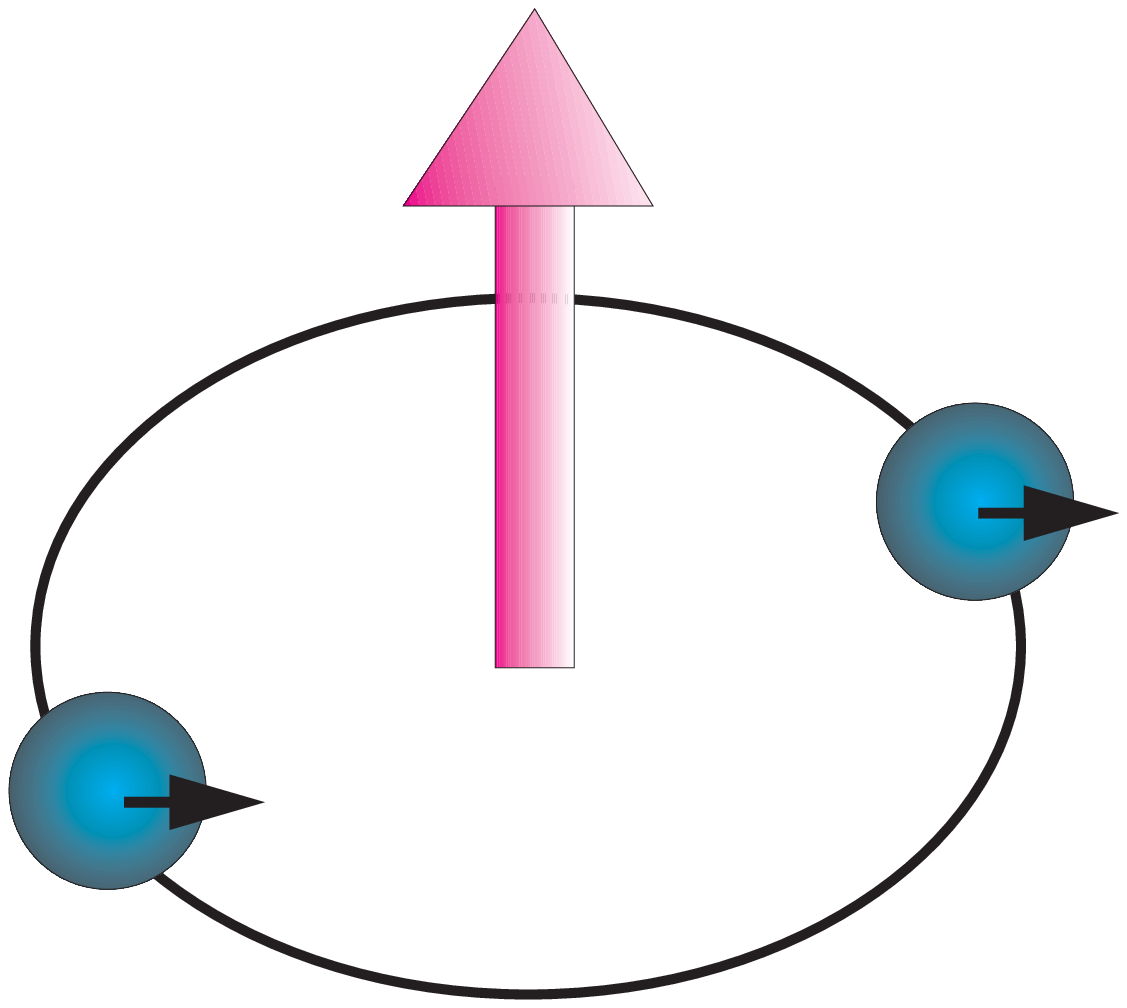,width=7.0cm,angle=0}}
\caption{The structure of the order parameter. The big arrow
corresponds to the orbital momentum $l$ of the Cooper pair, while
small arrows show the
 spins of the
electrons which add to $s=1$. In the frame in which $l_z=1$ one has $s_z=0$.}
\end{figure}

Below, we shall present in Sec. 2  the model and our approach. In
Sec. 3 we present the results of calculations, and finally we end
with our conclusions.

\section{The model and the approach.}
Our model of superconductivity in strontium ruthenate is
motivated by the experimental facts, summarised above. The
electronic structure features three Fermi surfaces, and all of
them are gaped to ensure the vanishing of the specific heat at
$T=0K$. Further, there  exists a single superconducting transition
and the gap has to both break the time reversal symmetry and  vanish
on the lines of the Fermi surface. We shall fulfill all these
constraints in the context of a realistic three dimensional band
structure with parameters fitted to the known details of the Fermi
surface,  and assumed effective attractive interactions between
electrons occupying various orbitals.

We thus take the following simple multi-band
attractive Hubbard Hamiltonian:
\begin{eqnarray}
  \hat{H}& =& \sum_{ijmm',\sigma}
\left( (\varepsilon_m  - \mu)\delta_{ij}\delta_{mm'}
 - t_{mm'}(ij) \right) \hat{c}^+_{im\sigma}\hat{c}_{jm'\sigma} \nonumber \\
&& - \frac{1}{2} \sum_{ijmm'\sigma\sigma'} U_{mm'}^{\sigma\sigma'}(ij)
 \hat{n}_{im\sigma}\hat{n}_{jm'\sigma'} \label{hubbard}
\end{eqnarray}
as our model which is to describe superconductivity in
Sr$_2$RuO$_4$. Here $m$ and $m^{\prime }$ refer to the three
Ruthenium $t_{2g}$ orbitals to be denoted $a=d_{xz}$, $b=d_{yz}$ and
$c=d_{xy}$ in the following. $i$ and $j$ label the sites of a body
centered tetragonal lattice. The hopping integrals $t_{mm^{\prime
}}(ij)$ and site energies $\varepsilon _{m}$ were fitted to
reproduce the experimentally determined three-dimensional Fermi
Surface~\cite{mac,berg}. We found that the set {$t_{mm'}$} shown
in Table 2 gave a good account of the data.

\begin{table}
\caption{\label{tabletwo}
 The values of parameters used in calculations.
  Here $t,t'$ are
 in-plane hopping integrals between $c$ ($d_{xy}$) orbitals,
 $t_{ax}, t_{bx}$ are in-plane hopping integrals between $a$
 ($d_{xz}$) orbitals along $\hat{\bf x}$ and $\hat{\bf y}$
 directions, $t_{ab}$ is the in-plane hopping between a and b orbitals
 along $\hat{\bf x}+\hat{\bf y}$, and $t_{hybr}$ and $t_{\perp}$ are out of plane
 hopping integrals along the bct body-centre vector from $a(b)$ to $c$ and from  $a(b)$ to $b(a)$,
 respectively.}
 \vspace{0.3cm} \hspace{-0.5cm} ~~
\begin{small}
\begin{tabular}{|c|c|c|c|c|c|c|c|c|c|c|c|c|}
 \hline
 & $t$ & $-t'$ & $t_{ax}$ & $t_{bx}$ & $t_{ab}$ & $t_{hybr}$
 & $-t_{\perp} $ & $-\varepsilon_{a}$ & $-\varepsilon_{b}$ & $-\varepsilon_{c}$
 & $U_{\parallel}$ & $U_{\perp}$ \\
 \hline
 meV  & $81.6$ & $36.7$ & $109.4$ & $6.6$ & $8.8$ & $1.1$
 & $0.3$ & $116.2$ & $116.2$ & $131.8$
 & $40.3$ & $48.2$ \\
 \hline
 \end{tabular}
 \end{small}
 \end{table}

In choosing the interaction parameters $U^{\sigma \sigma'}_{mm'}$
we adopted a frankly semi phenomenological approach. Namely, we
eschewed any effort to derive these from an assumed physical
mechanism of pairing and endeavoured to find a minimal, hopefully
unique, set of parameters which describes the available data. The
point of such strategy is that at the end we can claim to have
identified the position and orbital dependence of the interaction
and therefore provided guidance for constructing microscopic
models for specific mechanisms.

In the above spirit we consider two sets of interaction constants:
$U^\parallel_{mm'}$ for nearest neighbours in the plane and
$U^{\perp}_{mm'}$ for nearest-neighbour Ru-atoms  in the adjacent
plane  as indicated in Fig. 2.
 To limit further
the parameter space we need to explore, we assume that
$U^{cc}_{\parallel} \equiv U_{\parallel}$, $U_{\perp}^{aa} =
U_{\perp}^{bb} \equiv U_{\perp}$ and contemplate only
\begin{equation}
 U^\parallel_{mm'} =
\left(\begin{array}{ccc}
      u & u & u \\
      u & u & u \\
      u & u & U^\parallel
      \end{array}
      \right)~~~~~~{\rm and}~~~~~~
     U^\perp_{mm'}  =
          \left(
          \begin{array}{ccc}
           U_{\perp} & U_{\perp} & u' \\
           U_{\perp} & U_{\perp} & u' \\
           u'& u' & u'
         \end{array}\right)
\end{equation}
In fact, we take $u = u' = 0$ for most of our calculations and we
use only one in plane, $U_\parallel$, and one out of plane
$U_\perp$, parameters to fit  the experimental data on the
specific heat, $c_v(T)$, superfluid density $n_s(T)$ and the
thermal conductivity $\kappa_T(T)$. The purpose of the few
calculations with $u \neq 0$, $u' \neq 0$ was only to investigate
the stability of the $U_\parallel, U_\perp$ results to variations
in $U^\parallel_{mm'}$ and $U^\perp_{mm'}$.

Within the above model, we solved  the Bogolubov-de Gennes
equations:
\begin{equation}
 \sum_{jm'\sigma'}
 \left(
  \begin{array}{ll}
  E^\nu - H_{m,m'}(ij) & \Delta^{\sigma\sigma'}_{m,m'}(ij) \\
  \Delta^{*\sigma\sigma'}_{mm'}(ij) & E^\nu + H_{mm'}(ij)
  \end{array}
  \right)
  \left(
  \begin{array}{l}
   u^\nu_{jm\sigma'}\\
   v^\nu_{jm'\sigma'}
 \end{array}\right) = 0
\end{equation}
together with the self-consistency condition
\begin{equation}
\Delta^{\sigma\sigma'}_{mm'} = U^{\sigma\sigma'}_{mm'}(ij)
 \chi^{\sigma\sigma}_{mm'} (ij);~~~~~
\chi^{\sigma\sigma'}_{mm'}(ij) = \sum_{\nu} u^\nu_{im\sigma}
v^{*\nu}_{jm'\sigma'}(1 - 2f(E^\nu)),
\end{equation}
which follow from Eq. (1) on making the  usual BCS-like mean field
approximation\cite{ketterson}. Since $H_{mm'}(ij)$ and
$\Delta^{\sigma\sigma'}_{m,m'}(ij)$ depend only on the difference
$\vec R_i - \vec R_j$ the solution of Eq. (3) is rendered
tractable by taking its lattice Fourier transforms and, thus
transforming the problem into that of a
 12 $\times$ 12  matrix, eigenvalue problem
at each ${\bf k}$ in the appropriate Brillouin Zone.

For $u = u' = 0$ the general structure of
$\Delta^{\uparrow\downarrow}_{mm'}(\vec k)$
turns out to be of the form
\begin{eqnarray}
&& \Delta_{cc}(\vec k) = \Delta^x_{cc} \sin k_x +
\Delta^y_{cc} \sin k_y  \nonumber\\
&& \Delta_{mm'}(\vec k) = \Delta^z_{mm'}\sin {k_z c\over 2} \cos {k_x\over 2}
\cos{k_y \over 2} + \Delta^f_{mm'} \sin {k_x \over 2} \sin {k_y \over 2}
\sin{k_z c\over 2} \nonumber
\\
&& + \left( \Delta^x_{mm'} \sin {k_x\over 2} \cos {k_y\over 2}
+ \Delta^y_{mm'} \cos{k_y \over 2} \sin {k_y\over 2}\right)
  \cos{k_z c\over 2}
\end{eqnarray}
for $m,m' = a$ or $b$.

Our self-consistency procedure always converged on the amplitudes
$\Delta^x_{cc},  \Delta^y_{cc}$, $\Delta^z_{m,m'}(T)$,
$\Delta^f_{mm'}(T)$,  $\Delta^x_{m,m'}(T)$ and
$\Delta^y_{m,m'}(T)$ to an accuracy higher than 10$^{-4}$\%. The
principle results of such calculations are these amplitudes and
the corresponding quasiparticle energy eigenvalues $E_\nu(\vec
k)$. These eigenvalues are shown in Fig. 3,
 for the interaction values $U_\parallel$, $U_\perp$ given in Table 2, and
 with the constraint
that $\Delta^f_{m,n}(T)=0$. Evidently from the figure, on the
$\gamma$ sheet there is an absolute gap below $E_{\gamma,min}(\vec
k_F)$, in the quasi-particle spectrum, while the $\beta$ sheet is
gapless with a line of nodes in the gap functions at $k_z =
\pm{\pi\over c}$ This dramatically different gap structure and
symmetry on different sheets of the Fermi Surface is the striking
new results of an interaction matrix $U^{\sigma\sigma'}_{mn'}(ij)$
which couples electrons in different Ru-planes.

From the results in Fig. 3, one is encouraged to investigate this
form of interaction further because, on the one hand, the line of
zero gap on the  $\beta$ sheet explains the power-law behaviour of
various thermodynamic quantities and, on the other, on the fully
gapped $\gamma$ sheet $\Delta_{\gamma\gamma}(k) = \Delta_{cc}(k) =
\Delta^x_{cc}(\sin k_x a + i\sin k_y b)$ which implies the broken
time reversal symmetry demanded by a number of other experiments.
Indeed, for the above pair of interactions, $U_\perp$ and
$U_\parallel$, and the corresponding quasiparticle spectra we find
the specific heat $c_v(T)$ shown in Fig. 4 as a function of $T$.
Although these parameters were chosen to fit the experimental data
which is also shown, the agreement between theory and experiment
is truly remarkable. The point to appreciate is that in general
two different $U_\parallel$ and $U_\perp$ imply two separate
transitions at $T^\parallel_c$ and $T^\perp_c$. Thus, to agree
with the experiments which features a single transition at $T_c =
1.5$~K we had to use both degrees of freedoms. Namely, fitting
$T_c$ determined both coupling constants $U_\perp$ and
$U_\parallel$. Now, one might suggest that it is a short-coming of
our model that we have to rely on such an accidental coincidence
of $U_\parallel$ and $U_\perp$ to fit one number $T^{\rm
exp}_{c}$. However, in the light of the very good fit to the very
non trivial experimental variation of $c_v$ with temperature this
is not a strong objection. Indeed, the fact that having fitted to
$T^{\rm exp}_c$ only and we have reproduced the slope as $T$ goes
to zero and the size of the jump at $T_c$ has to be regarded as a
confirmation of our model.

\section{Concluding remarks.} 
Our calculations \cite{annett02} of the superfluid
density $n_s(T)$ \cite{DalevH} and thermal
 conductivity $\kappa(T)$ \cite{thermal-cond}, for
the same parameters as above, provide further
evidence in support of our model. Note that these calculations do
not involve additional adjustable parameters and thus the results
are not only qualitative consequences of nodes of the gap on the
$\alpha$ and $\beta$ sheets of the Fermi Surface but are also
quantitative predictions of the theory.

In what follows we conclude this brief survey of our very
encouraging results by two further important comments. The first
concerns the constraints $\Delta^f_{mm'}(T) = 0$ imposed at each
step of the selfconsistency cycle during the above calculations.
Without this constraint there would be a second transition $T^f_c
< T_c$, where $\Delta^f_{m,m'}(T)$ becomes non zero and this would
imply a peak in $c_v(T)$ at $T^f_c \simeq 0.2$~K, which has not
been seen experimentally. Fortunately, we have been able to show
by explicit quantitative calculations that a small amount of
disorder will eliminate the $ \Delta^f_{mm'}$ component of the
order parameters without changing the other amplitudes in Eq. (4)
very much. The details of these calculations will be published
elsewhere \cite{annett02}. This justifies, at the present level of
the mean-free path in samples on which the experiments were made,
the simultaneous neglect of disorder and the $f$-component
$\Delta^f_{m,m'}(T)$.

The other comment is that we have studied the stability of the
above model (Eqs. 1-5, Fig. 5) to introducing further interaction 
constants and found
it to be satisfactorily robust. In particular we have carried a
number of calculations with $u\neq 0$ and $u' \neq 0$ and, as we
shall report in a separate publication, we found \cite{annett02}
that the overall picture presented above remained the same.
Namely, our conclusion that a model with just two interaction
constants, one in plane $U_\parallel$, and one out of plane
$U_\perp$ (Fig. 4) roughly of the same size, around 50~meV, is capable of
explaining all the available data we have analysed remained valid.

\section*{Acknowledgments}
This work has been partially supported by KBN grant No. 2P03B 106 18 and the
Royal Society Joint Project.

\end{document}